%% file: RR-7542.tex
\documentclass[a4paper]{article}
\usepackage{RR}
\usepackage{hyperref}
\usepackage{mathptmx}
\usepackage{graphicx}
\usepackage{times}

\usepackage{multirow}

\RRdate{Janvier 2011}

\usepackage{color}
\definecolor{darkblue}{rgb}{0, 0, 0.45}
\definecolor{grey}{rgb}{0.5, 0.5, 0.5}
\definecolor{orange}{rgb}{1, 0.5, 0}

\RRtitle{\textsc{Portolan} : un Environnement de Cartographie Dirig\'ee par
les Mod\`eles}
\RRetitle{\textsc{Portolan}: a Model-Driven Cartography Framework}
\titlehead{Model-Driven Cartography} 
\RRauthor{Vincent Mah\'e
	\and Salvador Martinez Perez
	\and Hugo Bruneli\`ere
	\and Guillaume Doux
	\and Jordi Cabot}
\authorhead{V. Mah\'e \&
			S. Perez \&
			H. Bruneli\`ere\&
			G. Doux \&
			J. Cabot}
\RRresume{Traiter de grands volumes de donn\'ees pour en extraire des
informations utiles est une t\^ache essentielle au sein des entreprises. Pour
faciliter cette t\^ache, les techniques de visualisation ont \'et\'e
commun\'ement utilis\'ees, du fait de leur capacit\'e \`a pr\'esenter les
donn\'ees dans des vues synth\'etiques, plus faciles \`a comprendre et \`a
g\'erer.

Cependant, obtenir la bonne repr\'esentation visuelle pour un ensemble de
donn\'ees est un processus complexe de cartographie qui implique de nombreuses
\'etapes de transformation pour adapter le format des donn\'ees du domaine
au format des donn\'ees de visualisation attendu par les outils de
visualisation. Pour maximiser les avantages de la visualisation, nous proposons
\textsc{Portolan}, un environnement g\'en\'erique de cartographie dirig\'ee par
les mod\`eles qui facilite la d\'ecouverte des donn\'ees \`a visualiser, la
sp\'ecification des d\'efinitions de vues sur ces donn\'ees, et les
transformations pour franchir la distance avec les outils de visualisation.
Notre approche repose sur les composants Eclipse EMF Modeling et a \'et\'e
valid\'ee sur trois diff\'erents cas d'utilisation.}

\RRabstract{
	Processing large amounts of data to extract useful information is an essential
	task within companies. To help in this task, visualization techniques have
	been commonly used due to their capacity to present data in synthesized views,
	easier to understand and manage. 
	
	However, achieving the right visualization display for a data set is a complex
	cartography process that involves several transformation steps to adapt the
	(domain) data to the (visualization) data format expected by visualization
	tools. To maximize the benefits of visualization we propose Portolan, a
	generic model-driven cartography framework that facilitates the discovery of
 	the data to visualize, the specification of view definitions for that data
 	and the transformations to bridge the gap with the visualization tools. Our
 	approach has been implemented on top of the Eclipse EMF modeling framework
 	and validated on three different use cases. }
\RRmotcle{cartographie, visualisation, mod\`ele, IDM}
\RRkeyword{cartography, visualization, model, MDE}
\RRprojets{AtlanMod}

\RRtheme{\THCom}
\URRennes  

\begin{document}
\RRNo{7542}
\makeRR

\input{src_intro_new.tex}

\input{src_approach.tex}
\input{src_MD_visualization.tex}
\input{src_MDC_process.tex}
\input{src_tool.tex}

\input{src_experiment.tex}

\input{src_state_of_art.tex}
\input{src_discussion.tex}

\textbf{Acknowledgements}

The present work is being supported by the French ANR IDM++ project.

\bibliographystyle{abbrv}
\bibliography{Biblio}

\end{document}

%% file: src_intro_new.tex
\section{Introduction}


Business data is a key asset in any company. Efficient analysis and
understanding of this data can bring a significant competitive advantage.
Visualization techniques have been largely used for this purpose since the
beginning of computer science, from the early \textsc{apE} dataflow
toolkit\cite{dyer1990visualization}, the \textsc{Hy+} visualization
system\cite{consens1994architecture},
\textsc{AVS}\cite{upson2002application}, to the \textsc{InfoVis}
toolkit\cite{fekete2005infovis}.

Nevertheless, good data visualizations are not trivial to obtain. Visualization
itself is a complex process with its own techniques, formats and tools which
are different from those used in the business domain. Therefore, business data
must be processed (e.g. to select the subset we want to visualize) and then
transformed to the format expected by a given visualization tool.

This process, from raw data to a useful visualization, is a
complex process which needs to be managed in itself. It includes several
intermediate steps in order to get appropriate data, translate them to other
formats, merge them, bind them to corresponding visual items, and give
users an appropriate interface to \textit{play }with the data shown in the
views. Until now, this process has been generally created in an hard-coded and
adhoc manner (i.e. only useful for a given visualization tool and/or for a
specific business domain).

In this paper, we present an unified framework to deal with all these steps in a
more homogeneous way. We refer to this framework as a \textit{cartography}
framework\footnotemark[1]\footnotetext[1]{In the geographical domain, such a
process is named \emph{Map-Making}}. Moreover, our framework is based on the
use of Model-driven engineering techniques (MDE). MDE advocates the use of
models as key artefacts in all software engineering activities. Representing
the different components and manipulations in the cartography as models and
model transformations, we can manage all cartography elements in a more
homogeneous way. Therefore, the translation from raw business data to
visualization can be expressed as a set of model transformations.

In this sense, the contributions of this paper are threefold. First we define,
to the best of our knowledge, the first complete  Model-Driven Cartography
process. We have identified four high level tasks and the different kinds of
actors that are needed to bridge the gap between raw data and visualization.
Secondly, we predefine several of the (modeling) elements needed in the
process, as the cartography metamodel. Finally, we provide an extensible tool
support in the form of a  ready-to-use platform that can be easily tailored
to the needs of each given domain.

This paper is structured as follows: this introduction is followed by a
general presentation (section \ref{Over All section}) of our approach. The three
contributions are detailed in next sections: the Cartography Metamodel in
section \ref{MDV section}, the Model-Driven Cartography process in section
\ref{MDC Process section} and, in section \ref{Portolan section}, our
implementation, the \textsc{Portolan} platform. Three use cases illustrate the
different features of our approach in section \ref{Experiments section}.
Section \ref{Related Works section} compares with the related works, followed
by the conclusions and further work in section \ref{Discussion section}.

%% file: src_approach.tex
\section{Overall View of \emph{Model-Driven Cartography}}
\label{Over All section}

Our model-driven cartography process is built around a pivot cartography
metamodel that acts as an intermediate representation in the chain of
transformations needed to go from raw data to visualization. This intermediate
representation facilitates the resue of most of the cartography elements across
different domains. In this section, we introduce this metamodel, the process
around it and the tool support we provide. All these concepts are then
detailed in the next sections.

\begin{figure}\centering
  \includegraphics[width=\linewidth]{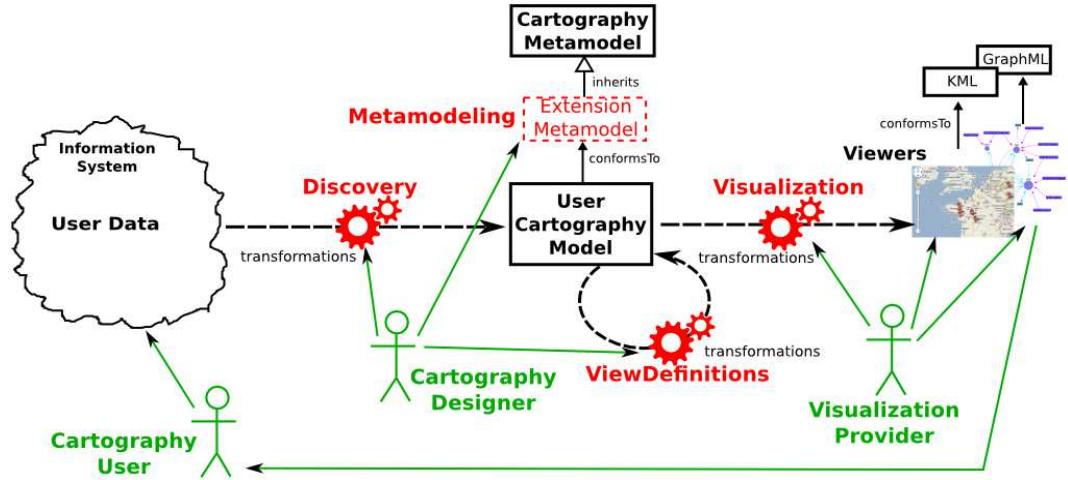}
  \caption{Model-Driven Cartography}
  \label{Figure: cartography-schema}
\end{figure}

	\subsection{Cartography Metamodel}

To deal with the data-to-visualization issues, we rely on a model-driven
mechanism. A dedicated metamodel is thus needed to provide a uniform
representation for the data to be visualized. We have designed an  all-purposes
Cartography Metamodel (on top of Fig.\ref{Figure: cartography-schema}) with
generic types (as Entity or Relationship) useful for visualization purposes.
The user's data is then expressed in an User Cartography Model (center of the
schema) which conform to this (possibly extended) metamodel.

After, this data is used as input of several viewers (right side of
Fig.\ref{Figure: cartography-schema}) able to read data conforming to this
cartography metamodel and transform it into visual items (nodes, edges,etc).

	\subsection{Cartography Process}
	\label{Cartography Process}
	
A cartography process can be summarized in four high-level tasks (built around
the previous cartography metamodel) required in the data-to-visualization
process for any given domain:
	\begin{enumerate}
	  \item \emph{Metamodeling}: capture the structure of the domain concepts to
	  be visualized. If necessary, the cartography metamodel can be extended to
	  tailor it to the concepts of the input domain;
	  \item \emph{Discovery}: inject user data into a central cartography model
	  which conforms to the domain extension of the cartography metamodel;
	  \item \emph{View Definitions}: filter the data in the user cartography model
	  to get partial and/or computed views of it to be visualized;
	  \item \emph{Visualization}: obtain graphical displays of user's data as
	  defined in the previous views.
	\end{enumerate}

As part of the process definition, we have identified three different kinds of
actors that should take part in it:
	\begin{itemize}
	  \item \emph{Cartography User}: he is the person the whole cartography is
	  designed for; he provides its \textbf{data} as input of the process and use
	  the corresponding \textbf{viewers} to visualize and take decisions from
	  the generated views; he does not need skills in MDE nor in Cartography;
	  \item \emph{Cartography Designer}: he is an engineer with enough skills in
	  MDE to be able to specify the \textbf{domain metamodel} extension, to
	  chain and use \textbf{discovery} tools (generating the user central
	  cartography model) and to write useful \textbf{view definitions}; 
	  \item \emph{Visualization Provider}: it can be an external company or
	  organism, which builds pluggable model transformations and
	  \textbf{visualization} tools relying on the cartography metamodel; the
	  visualization may be provided as a component-on-the-shelf.
	\end{itemize}

The data translations needed in our cartography approach rely mainly on
the Model-Driven Interoperability principle, as proposed by B\'ezivin et
al.\cite{bezivin2005model}. This is a major benefit of using MDE for
Cartography.

	\subsection{Tool Support}
	
The characterization of these four high level tasks, with the identification
of the three kinds of users, has driven to an effective implementation of our
approach in a Model-Driven Cartography platform. The developed tooling is based
on Eclipse and available as an open-source application\footnotemark[6].

This platform  provides MDE components related to the visualization framework,
and ready-to-use tools which handle most parts of the Cartography process. They
can be extended and tuned to user's specific needs.

%% file: src_MD_visualization.tex
\section{Cartography Metamodel}
\label{MDV section}

\begin{figure}\centering
  \includegraphics[width=\linewidth]{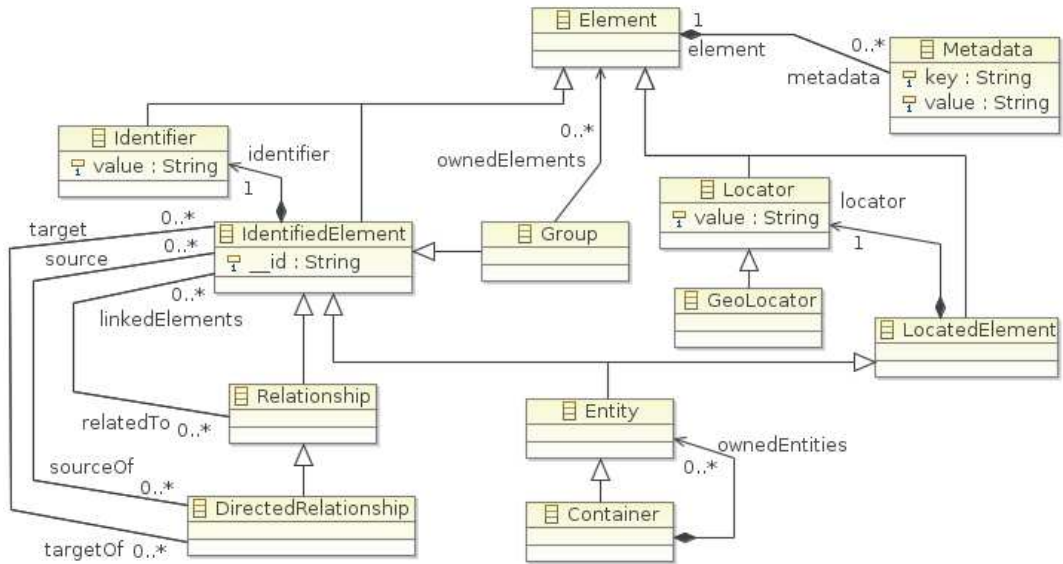}
  \caption{Cartography generic metamodel}
  \label{Figure: cartographymm}
\end{figure}

Our cartography metamodel provides an intermediate and generic representation
for all kinds of domain. The cartography metamodel contains all important
concepts needed to achieve a proper data visualization. This core metamodel is
presented in Fig. \ref{Figure: cartographymm}. Table \ref{metamodel-table}
lists the main concepts of the metamodel (right column) and the visualization
requirements that motivate their presence (left column).

This metamodel can be used as-is, but users can also extend it to adapt it to
the important concepts in their domain, by creating a kind of Domain Specific
Language (DSL) for cartography. However, we have to notice that, as the
visualization part relies only on the core metamodel types (with possible
identification of the effective type of each instance through reflexivity), it
is generic and domain-agnostic.

\begin{center}
 \begin{table}
  \begin{tabular}{p{8cm}p{0.3cm}p{3.5cm}}
  
    \hline
	  \textbf{Task and what it implies} & & \textbf{Corresponding Types}\\
    \hline
  
	 A view has to display a name for each node & & \textsf{IdentifiedElement} (or
	 \textsf{Entity}) and its \textsf{Identifier}\\
    \hline
	 
	 Maps, geographical or not, need to locate elements & & \textsf{LocatedElement}
	 (\textsf{Entity}), and \textsf{Locator} (\textsf{GeoLocator})\\
    \hline
	 
	 Graphs require linking related elements together & & \textsf{Relationship}\\
    \hline
	 
	 Some relations are one-way & & 
	 \textsf{DirectedRelationship}\\
    \hline
	 
	 A grouping $/$ containment notion is important to cluster data & &
	 \textsf{Group}, \textsf{Container}\\
    \hline
	 
	 Each item may have specific individual properties, implying a \emph{type
	 object}\cite{johnson1997type} mechanism & & \textsf{Metadata}\\
 
   \hline
  \end{tabular}
  \vspace{3ex}
 \caption{Visualization needs and corresponding metamodel types}
 \label{metamodel-table}
 \end{table}
\end{center}
\vspace{-8ex}

%% file: src_MDC_process.tex
\section{Model-Driven Cartography Process}
\label{MDC Process section}

We detail here the four high level tasks of a Model-driven Cartography process
that we introduced in section \ref{Cartography Process}: \emph{Metamodeling},
\emph{Discovery}, \emph{View Definitions}, and \emph{Visualization}.

	\subsection{Metamodeling of the domain}
		\label{Metamodeling}

The metamodeling step consists in the specification of the targeted user's
domain, as an extension of the cartography metamodel. Three cases are
relevant:
	\begin{itemize}
	  \item the Cartography core metamodel can be used as-is by customers, who
	  will then build the discovery tools and generate the central model using
	  only the core metamodel types (illustrated in Fig. \ref{Figure:
	  cartographymm});
	  \item the Cartography metamodel can be extended by an engineer (the
	  \emph{Cartography Designer} role) to add types closer to the user domain.
	  These new types are linked to the core ones through inheritance
	  relationships. In this case, the first steps in the metamodeling phase are
	  devoted to specifying the main concepts of the domain and deciding the base
	  class in the core metamodel for each of them. The resulting types of the
	  metamodel extension define a cartography DSL for the domain, enabling the
	  corresponding tooling, transformations and modeling of the domain. An
	  example of such a DSL is visible in Fig. \ref{Figure: BusinessCartography},
	  illustrating this modeling initial step on a domain which concerns software
	  tools.
	  \item an existing domain metamodel can also be reused as cartography
	  metamodel if the cartography designer is able to define inheritance
	  relationships with the core metamodel types. This also affects the relations
	  between elements. References must be reified in order to inherit from
	  (\textsf{Directed})\textsf{Relationship} core type; Fig. \ref{Figure:
	  EclipseMMs} illustrates this case.
	\end{itemize}

These three cases have been experimented (see section \ref{Experiments
section}), validating this metamodeling phase.

	\subsection{Discovery of the Data}
		\label{Discovery}

The goal of this step is to transform raw data from different sources (and
coming from different technical spaces\cite{kurtev2006model}), provided as-is by
the final user (the \emph{Cartography User} actor), into a central cartography
model which conforms to the (possibly extended) core cartography metamodel.
These tasks are handled by an engineer with skills in MDE (\emph{Cartography
Designer} role).

This phase is expressed as a sequence of text-to-model and model-to-model
transformations. A classical discovery chain is composed of the following steps:
\begin{enumerate}
  \item An injection process takes the raw data and transforms it into a
  model-based representation. For instance for XML-based raw data, a predefined
  XML injector express this data as an instance of the XML metamodel. At this
  stage, the structure of the data has not been changed, only the nature of its
  representation has been modified from XML to model. The main benefit of this
  step is that the data is, now, expressed as a model which we can directly
  use all MDE techniques on to manipulate this data.
  \item A first model-to-model transformation, depending on the nature of raw
  data format, takes the injected model and transforms it into a domain model.
  Following the previous example, in this step the information included in the
  XML model is extracted and represented as a domain model instance of an
  appropriate domain metamodel for the user's activity. E.g., if the XML-based
  raw data included information about personnel of a company, in this step,
  this data is expressed in terms of a
  HHRR\footnotemark[2]\footnotetext[2]{Human Health Risk Resources} metamodel.
  In short, this domain metamodel corresponds to the elements in the XML Schema
  for the raw data.
  \item The domain model then needs to be mapped to a cartography model. A
  model-to-model transformation (written by people with knowledge of both data
  domain and the core metamodel) takes care of this. The key task in this step
  is to select the most appropriate cartography concept for each domain element.
  \item If multiple sub-processes are involved, a merging transformation
  produces the final result of the Discovery task.
\end{enumerate}

\begin{figure}\centering
  \caption{Cartography: a discovery sub-process example}
  \includegraphics[width=\linewidth]{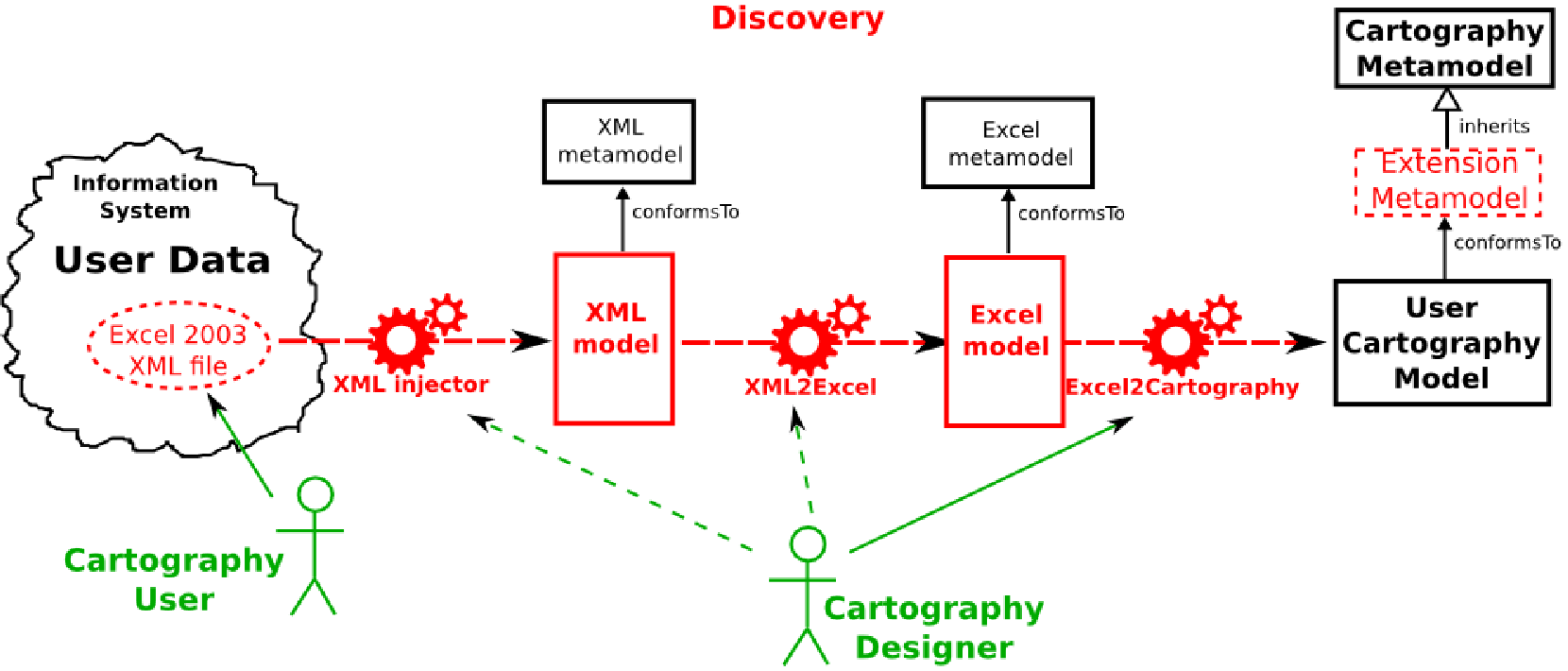}
  \caption{Cartography: a discovery sub-process example}
  \label{Figure: Step2}
\end{figure}

We illustrate such a sub-process on an example from an Excel data file to the
central model using an XML injector and two model transformations (Fig.
\ref{Figure: Step2}):
	\begin{enumerate}
	  \item The \emph{Cartography User} brings a data file in MS Excel 2003 XML
	  format;
	  \item The engineer (\emph{Cartography Designer}) runs a XML injector on this
	  XML textual file to generate the corresponding model (which then conforms to
	  the XML metamodel);
	  \item This engineer then processes the previous output XML model with an
	  existing XML2Excel transformation which explores the XML model and generates
	  the underlying Excel items, as a model which conforms to the Excel
	  metamodel (with \textsf{Worksheet}, \textsf{Table}, \textsf{Row},
	  \textsf{Column} types);
	  \item As the last step implies domain specific elements and structure, the
	  corresponding model-to-model transformation must be written by the
	  \emph{Cartography Designer}, with some knowledge of the content of the
	  Excel file; the transformation processes its rows (as elements of the model)
	  depending on their specific structure and cells and on the corresponding
	  types within the user's extension of the cartography metamodel;
	  \item As all the data were provided within the same Excel file, no additional
	  transformation is needed here in order to merge all generated cartography
	  models in one central model.
	\end{enumerate}

This Discovery phase reuses many MDE components which already exist (for
instance the ATL Zoo\footnotemark[3]\footnotetext[3]{ATL Transformation Zoo:
http://www.eclipse.org/m2m/atl/atlTransformations/} includes more than 200
transformations to exchange data between different formats).

	\subsection{Edition of the View Definitions}

View Definitions are transformations of the (extended) cartography model to
filter the data and/or compute derived information that must be visualized.
They are expressed as model-to-model transformations. The result is another
cartography model that still conforms to the same cartography metamodel. This
subset can then be processed as the original one by the provided
visualizations.

\begin{figure}\centering
  \includegraphics[width=300pt]{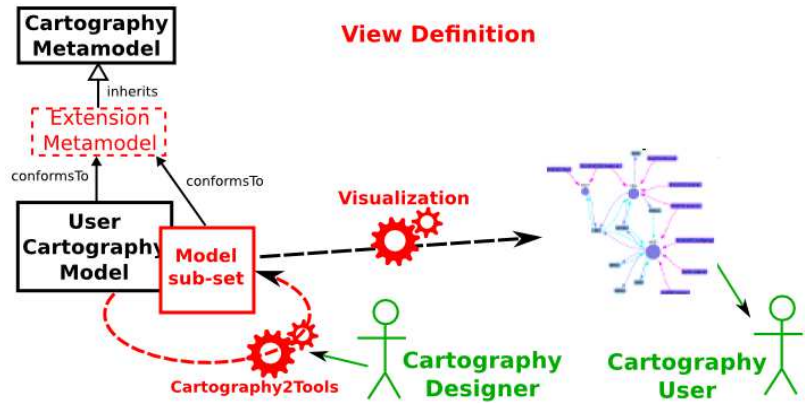}
  \caption{Cartography: an example of a view definition}
  \label{Figure: Step4}
\end{figure}

The principle of view definitions is illustrated (Fig. \ref{Figure: Step4}) on
an example relying on a DSL similar to Fig. \ref{Figure: BusinessCartography},
with a transformation which computes direct links between \textsf{Tool}
instances over the corresponding \textsf{Format} instances:
	\begin{enumerate}
	  \item An engineer (the \emph{Cartography Designer} role) writes a
	  Cartography2Tools transformation taking the central cartography model as
	  input and a model conforming to the same metamodel as output; it generates
	  only the instances of Tool type in the destination model and computes
	  relationships between tools by replacing Format instances and corresponding
	  links to tools by direct Export/Import instances;
	  \item The resulting subset model, as it relies on the same metamodel, but
	  with only \textsf{Tool} entities and \textsf{Export/Import} directed
	  relationships, can then be processed by provided visualizations as if it was
	  the central model;
	  \item The \emph{Cartography User} gets the corresponding display, which is a
	  filtered view (\textsf{Tool} nodes linked by \textsf{Export/Import} edges) of
	  the whole picture.
	\end{enumerate}

The main expectation for such edition work is to give engineers an access to
a model-driven workbench with the required environment for editing view
definitions. As view definitions rely on model-to-model transformations, a
transformation language environment (including an editor, outline, debugger,
compiler, etc) is mandatory in a dedicated Cartography environment.

	\subsection{Visualization of the Model}

Visualization components can be provided by external \emph{Visualization
Providers}. Each visualization relies on a specific viewer and a dedicated
transformation chain. This chain is a process opposite to the Discovery one, as
it takes a cartography model as input and generates raw (often textual) data
files which can be read natively by the targeted viewer:
\begin{enumerate}
  \item A model-to-model (view-oriented) transformation changes the
  cartography model into a visualization model which conforms to the viewer
  metamodel.
  \item Another model-to-model (data-oriented) transformation takes the viewer
  model and translates it to the targetted textual syntax as a model embedding
  the visualization items.
  \item A model-to-text transformation re-expresses this model into the
  textual description format expected by the viewer. This file is generated
  by an extractor component.
  \item The final textual file can be directly read by the viewer which
  displays the corresponding tree, graph, or other visualization. The
  \emph{Cartography User} can then manipulate the view in order to better
  understand and analyze the business data.
\end{enumerate}

Our tool (section \ref{Portolan section}) embeds the metamodels corresponding to
the formats needed by several viewers. We have used existing metamodels (like
GraphML, from a previous project\footnotemark[4]) and we defined the common
ones that were lacking (such as the KML one) to provide a ready-to-use
platform.

\begin{figure}\centering
  \includegraphics[width=\linewidth]{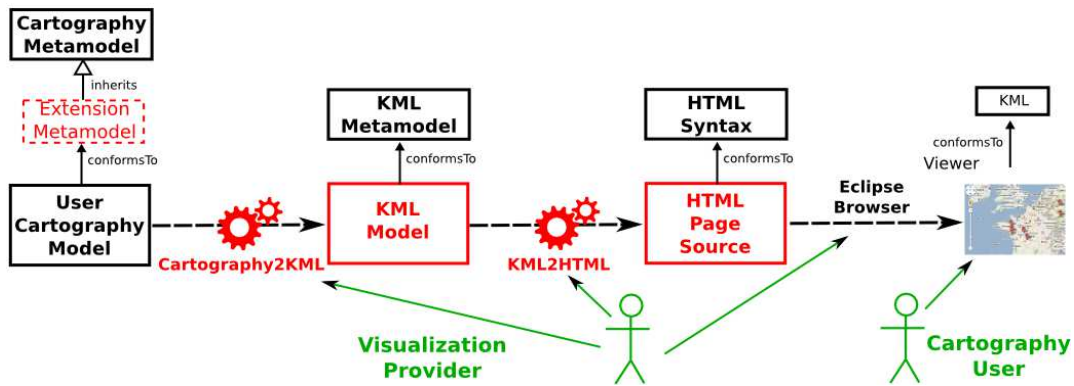}
  \caption{Cartography: one visualization process}
  \label{Figure: Step3}
\end{figure}

An example of such visualization process produces GoogleMaps visualization for
the geo-located elements of the user's cartography model (Fig. \ref{Figure:
Step3}):
	\begin{enumerate}
	  \item A model-to-model Cartography2KML transformation, written by the
	  \emph{Visualization Provider} of a GoogleMaps view, processes the central
	  cartography model and creates, for each \textsf{LocatedElement} instance
	  which has a \textsf{GeoLocator} kind as its \textsf{locator} attribute, an
	  element into a GoogleMaps model which itself conforms to the Keyhole Meta
	  Language (KML)\footnotemark[5]\footnotetext[5]{KML:
	  http://code.google.com/apis/kml/};
	  \item The output KML model is processed within an EMF-based program which
	  takes the KML items and generates an HTML page source with the corresponding
	  KML tags;
	  \item The HTML textual source corresponding to the KML model is passed to
	  the internet browser embedded within Eclipse;
	  \item The code of the HTML page calls the GoogleMaps viewer and pass the KML
	  data to it, which displays the user's items on the corresponding map.
	\end{enumerate}

The visualization component can be provided with internal mechanisms which hide
the whole chain, so the user only sees a button to launch the chain and
automatically gets the final result in the display.

%% file: src_tool.tex
\section{Tool Support of Model-Driven Cartography}
\label{Portolan section}

\begin{figure}\centering
  \includegraphics[width=150pt]{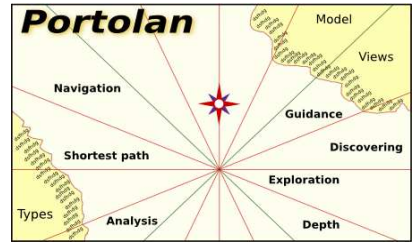}
  \caption{\textsc{Portolan}}
  \label{Figure: Portolan}
\end{figure}

As \emph{portolans} were first nautical charts to help navigation in $14^{th}$
century, the \textsc{Portolan} application (Fig. \ref{Figure: Portolan})
aims to help decision-makers to explore, discover, navigate, analyze the
collections of data of their company.

The \textsc{Portolan} Cartography Framework\footnotemark[6]
\footnotetext[6]{\textsc{Portolan}: http://code.google.com/a/eclipselabs.org/p/portolan/}
is our implementation of the Model-Driven Cartography approach presented in this
report. In order to be fully usable as an implementation, the platform provides
default generic components for each of the four steps identified in our
approach, providing a ready-to-use Cartography toolbox. As an effective
platform, it demonstrates the feasibility of our conceptual proposal.

The next sections detail the main features and underlying architecture of the
\textsc{Portolan} platform.

	\subsection{Overview}

The \textsc{Portolan} prototype embeds two main perspectives (in the Eclipse
sense), each of them addressing distinct sets of users:

\begin{enumerate}
  \item the \textbf{Decision-Maker} perspective (Fig. \ref{Figure:
  Decision-maker Perspective}), centered on visualization aspects, allows
  graphical display, navigation and manipulation of the user's data; the goal
  is to efficiently help the decision-maker to better understand the situation,
  and then take appropriate decisions; the decision-maker perspective can be
  considered as the end-user application offered by the prototype; it targets
  the \emph{Cartography User} actor early defined in our approach; this
  perspective hides the cartography process, so its user only sees the final set
  of available visualizations;
  
  \item the \textbf{Engineer} perspective (Fig. \ref{Figure: Engineer
  Perspective}) is centered on discovery work and definition of views; it
  allows to define and customize transformations, according to specified
  criteria and constraints, and to generate the central cartography model to be
  visualized by the decision-maker; the goal is to provide a
  development-oriented environment for the cartography process; it is a
  workbench integrating the ATL\footnotemark[7]\footnotetext[7]{Eclipse M2M
  AtlanMod Transformation Language (ATL) Project:
  {http://www.eclipse.org/atl/}} perspective (for implementing the needed
  model-to-model transformations) together with multiple DSLs and projectors
  for standards like KML, GraphML, etc ; this perspective is intended to be
  used by the \emph{Cartography Designer}, but the \emph{Visualization
  Provider} may also benefit from it to develop visualization extensions; this
  perspective handles all the cartography process.
\end{enumerate}

\begin{figure}\centering
  \includegraphics[width=\linewidth]{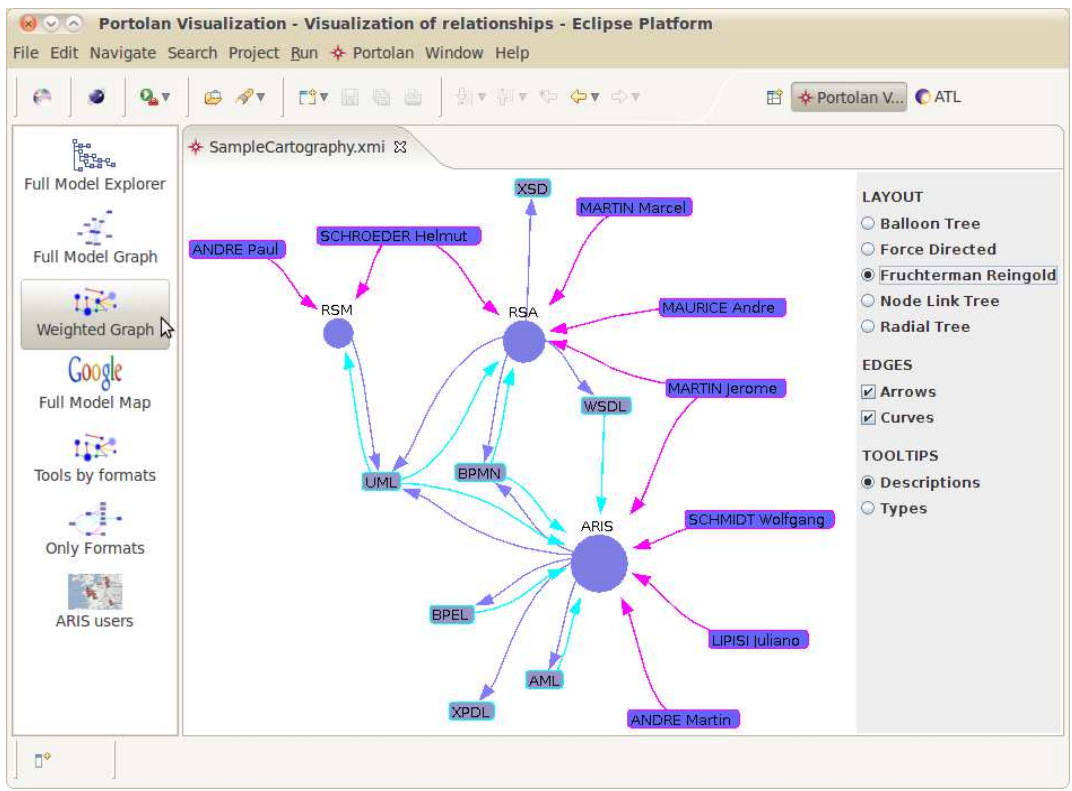}
  \caption{\textsc{Portolan}: the Decision-maker perspective}
  \label{Figure: Decision-maker Perspective}
\end{figure}

\begin{figure}\centering
  \includegraphics[width=\linewidth]{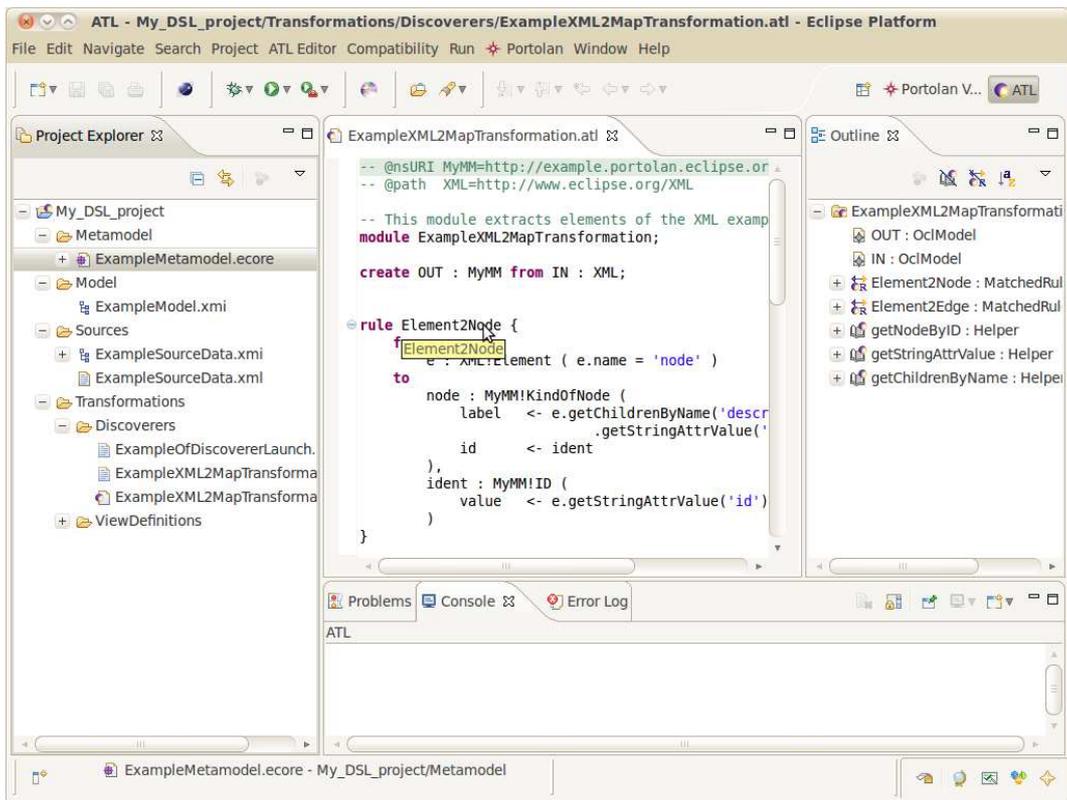}
  \caption{\textsc{Portolan}: the Engineer perspective}
  \label{Figure: Engineer Perspective}
\end{figure}

	\subsection{Underlying Architecture}

\begin{figure}\centering
  \includegraphics[width=300pt]{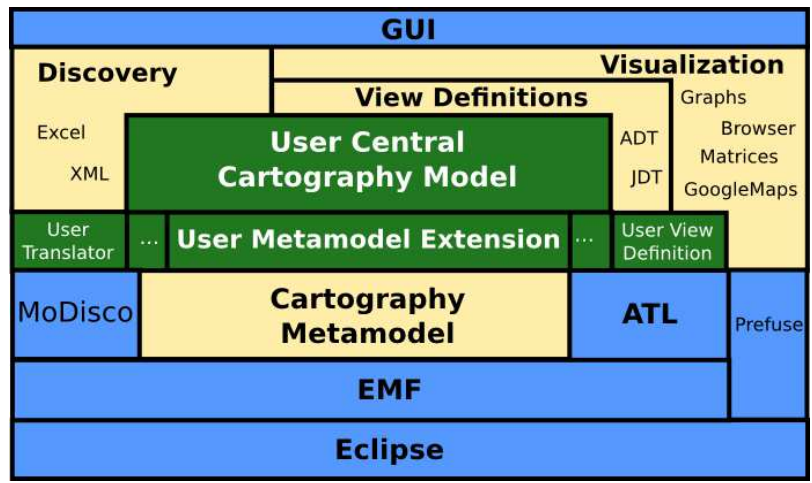}
  \caption{\textsc{Portolan} architecture}
  \label{Figure: Portolan_Architecture}
\end{figure}

The \textsc{Portolan} cartography platform includes multiple model-driven components
on top of Eclipse, as presented in Fig. \ref{Figure: Portolan_Architecture}.

		\subsubsection{Eclipse platform}

The lower elements in the schema are the part of \textsc{Portolan} which relies
on existing Eclipse components, with an emphasis on the Eclipse Modeling
project: EMF for the model handling, ATL\cite{jouault2008atl} for model
transformation, MoDisco\cite{madiot2010modisco} for model discovery, edition
and navigation capabilities, etc.

As in any Eclipse tooling, the developments are realized in a workspace so that
is possible to directly benefit from the standard Eclipse tooling such as
available versioning systems (e.g. SVN or CVS), task managers (e.g. Mylin),
etc.
    
		\subsubsection{Portolan core}
		
The upper elements of the schema are the cartography framework high-level
components. They are related to the four steps presented earlier:
\emph{Metamodeling}, \emph{Discovery}, \emph{View Definitions} and
\emph{Visualization}.

This part includes the extension mechanisms presented in subsection
\ref{Extension section}. An Eclipse-based graphical user interface gives
useful handles on both components.

		\subsubsection{User's parts}

The intermediate elements are the components which are customer-specific. The
company may adapt or extend the provided sample metamodel and transformations
in order to specialize them to its own domain of cartography. The user's
metamodel extension gives the abstract syntax for the user's central
cartography model. Translators are transformations used in discovery chains.
View definitions are transformations to compute the user's model into subsets
or metrics to be displayed in the visualization perspective.

The platform provides default elements for this user's parts (giving a
ready-to-use application), which the company may adapt or replace.

Most of the features provided by the \textsc{Portolan} cartography platform are
related to one of the four leading tasks highlighted in our Model-Driven
Cartography approach (see Fig. \ref{Figure: cartography-schema}):
\begin{itemize}
  \item \emph{Metamodeling}: PortolanCore cartography metamodel, Ecore modeler;
  \item \emph{Discovery}: ATL XML injector, Excel \& XML metamodels;
  \item \emph{View Definition}: ATL development environment;
  \item \emph{Visualization}: ATL XML extractor, GraphML and KML metamodels,
  predefined viewers.
\end{itemize}

	\subsection{Predefined Viewers}

The Portolan Cartography Framework is shipped with predefined functional
viewers:
\begin{itemize}
  \item a Prefuse-based\cite{heer2005prefuse} graph viewer: it relies on
  \textsf{Entity} and \textsf{Relationship} types of the Cartography metamodel
  to display corresponding graphs, offering multiple layouts (force directed,
  nodes tree, LinLog\cite{noack2007energy}, etc);
  \item a GoogleMaps view: this viewer displays on a GoogleMaps page the data
  items which have geolocated information, relying on the \textsf{GeoLocator}
  type.
  \item the MoDisco\cite{madiot2010modisco} model browser: this tool
  facilitates the navigation of the central model based on the type of the
  elements
  
\end{itemize}

	\subsection{Extension Mechanisms}
	\label{Extension section}

The \textsc{Portolan} cartography platform integrates all the provided features
within extension mechanisms in order to apply each feature to different domains
and be able to plug in \textsc{Portolan} other viewers and cartography tools.

		\subsubsection{Metamodel Extension}

The cartography metamodel presented in our approach (see section
\ref{Metamodeling} and Fig. \ref{Figure: cartographymm}) is embedded in
\textsc{Portolan} as the \emph{PortolanCore} Ecore metamodel. A company's
engineer (acting as \emph{Cartography Designer}) can extend this metamodel with
types closely related to its business domain by creating an Ecore diagram with
the provided Ecore tools and making these types inherit from PortolanCore types.
This metamodel extension only needs to be declared in the \textsc{Portolan}
preferences to be used as the reference cartography metamodel.

See section \ref{Experiments section} and Fig. \ref{Figure: EclipseMMs} for an
illustration of the way the core metamodel can be extended.

		\subsubsection{Generic Visualization Extension}

This extension point is the central registration mechanism of visual
functionalities inside \textsc{Portolan}. It must be used by every
\emph{Visualization Provider} to plug its viewer in order to make it available
to end-users. Some default viewers are provided with the \textsc{Portolan}
prototype (Prefuse-based graph viewer, Modisco model browser, GoogleMaps
viewer).

The technical implementation relies on Eclipse Extension Points mechanism. The
\emph{Visualization Provider} creates his viewer as an Eclipse editor, declares
it using the corresponding Editor Extension, and also declares it as a Portolan
Visualization extension, with respect to the corresponding schema. An example of
such a \textsc{Portolan} declaration is given here:
\begin{verbatim}
<extension
    point="fr.inria.portolan.visualization.fullView">
  <full_view
    commandId=
        "fr.inria.portolan.visualization.prefuse.commands.fullModel"
    editorId=
        "fr.inria.portolan.visualization.prefuse.ui.PrefuseEditor"
    iconPath="icons/prefuse_force.png"
    id="fr.inria.portolan.visualization.prefuse.view"
    name="Portolan Full Model Graph View Definition"
    pluginId="fr.inria.portolan.visualization.prefuse"
    text="Full Graph (Single)"
    tooltip="Multi relationships are managed multiple single edges">
  </full_view>
</extension>
\end{verbatim}

The underlying mechanism inserts the corresponding button into the view bar of
the Visualization perspective (as in the Decision-Maker perspective of Fig.
\ref{Figure: Decision-maker Perspective}), thus giving the \emph{Decision-maker}
a direct access to this new visualization.

		\subsubsection{End-User View Definition Extension}

In order to filter the central cartography model and get a subset of it, a
\emph{Cartography Engineer} may write ATL transformations which do computations
on this model and produce filtered models. The \textsc{Portolan} View
Definitions mechanism give the ability to declare such filters and get
corresponding button in the \textsc{Portolan} main tool bar (left side of the
Visualization perspective on Fig. \ref{Figure: Decision-maker Perspective}). When a user clicks on such a
View Definition button, the corresponding command is run: the transformation
applies on the central cartography model and the resulting model is passed to
the specified editor. The result is a visualization with only the filtered
elements.

As for visualizations, the \emph{View Definitions} mechanism relies on Eclipse
Extension Points. The \emph{Cartography Engineer} declares its ATL
transformation using a specific \textsf{*.vd} file with extension parameters as
illustrated in the next example. This \textsf{*.vd} file also specifies the
final Visualization editor which will be used to display the resulting subset
of the central cartography model.
\begin{verbatim}
<plugin>
  <extension
      point="fr.inria.portolan.visualization.viewDefinition">
    <view_definition
      id="ToolsCartography.example.ExtractTools"
      categoryId=
"fr.inria.portolan.visualization.prefuse.commands.viewDefinitionCategory"
      iconPath="Transformations/ViewDefinitions/ExtractTools.png"
      commandId="prefuse.tools.with.licenses.command"
      name="Tools with Licenses"
      text="View: Tools Licenses"
      transformationPath=
         "Transformations/ViewDefinitions/ExtractTools.asm"
      postName="_Tools"
      editorId=
         "fr.inria.portolan.visualization.prefuse.ui.PrefuseEditor"
      shortMM="Cartography">
    </view_definition>
  </extension>
</plugin>
\end{verbatim}

The \textsc{Portolan} Model-Driven Cartography Framework has been validated by
applying it to three use cases, covering the three distinct possible uses of the
metamodel identified in section \ref{Metamodeling}.

%% file: src_experiment.tex
\section{Experiments}
\label{Experiments section}

The \textsc{Portolan} cartography platform has its origin in the visualization
needs of several industrial projects with different requirements.

In what follows, we present three use cases that explore three potential ways to
use our framework by:
\begin{itemize}
  \item making a direct use of the cartography metamodel;
  \item extending the cartography metamodel with a domain-specific extension;
  \item plugging a previously existing DSL in the cartography metamodel.
\end{itemize}

\input{src_uc_inria.tex}

\input{src_uc_bnpp.tex}


\input{src_uc_eclipse.tex}

%% file: src_uc_inria.tex
	\subsection{Use of Core Metamodel: Collaborations within an Organization}
	
INRIA (National French Research Institution in Computer Science) wanted a
representation of its internal organization (areas, departments, human
resources) along with its research activity and results, specially regarding a
visual representation of the collaboration between research teams. Since this
collaboration can be simply represented as a graph, the predefined cartography
metamodel does not need to be extended in this scenario.

\begin{figure}\centering
  \includegraphics[width=\linewidth]{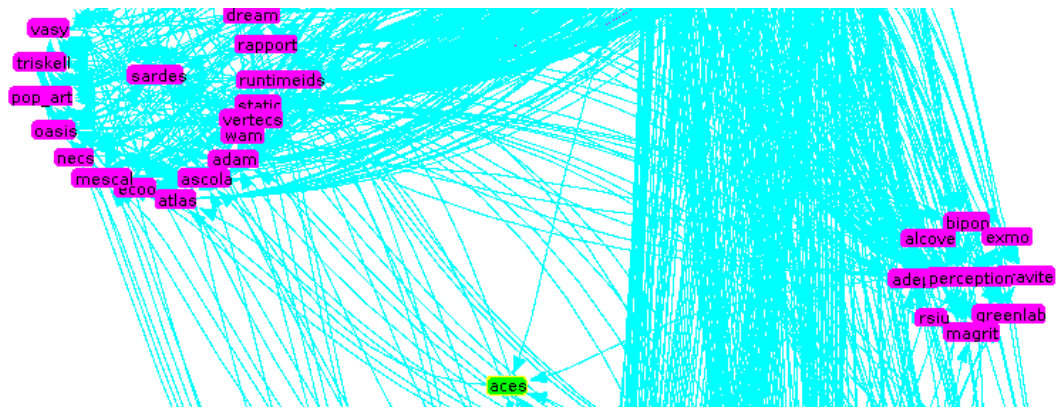}
  \caption{INRIA teams collaborations (excerpt)}
  \label{Figure: INRIA model}
\end{figure}

To populate the cartography metamodel we injected the XML documents describing
the collaboration into a specific collaboration metamodel. A model to model
transformation was in charge of translating this collaboration information in
terms of the cartography metamodel. From there, we can automatically visualize
the collaboration relationships. In Fig. \ref{Figure: INRIA model}, we can see a
sample of the resulting visualization.

%% file: src_uc_bnpp.tex
	\subsection{Core Metamodel Extension: Map of Business Tools and Formats}

A large bank was interested in visualizing the possible interoperability (in
terms of compatibility of the import/export formats) scenarios between the huge
number of tools used in its different business units.

\begin{figure}\centering
  \includegraphics[width=\linewidth]{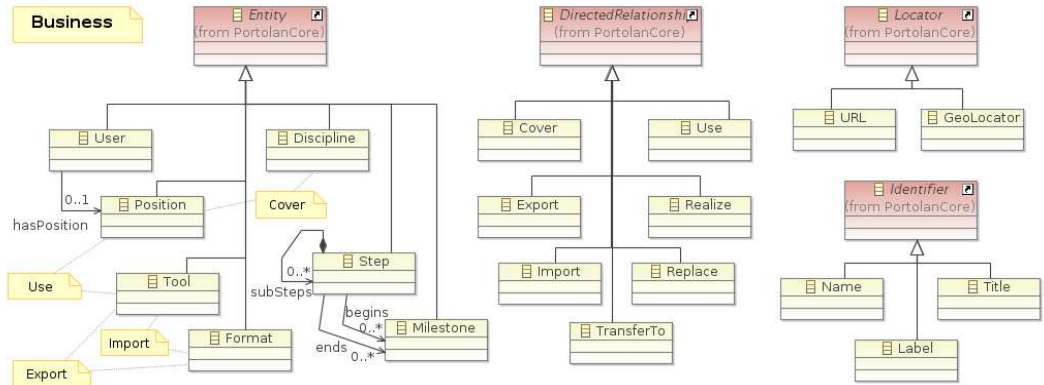}
  \caption{Business Cartography metamodel}
  \label{Figure: BusinessCartography}
\end{figure}

According to the requirements, we extended the cartography metamodel with new
Tool and Format entities together with Import and Export relationships. Some
additional concepts (like User or Milestone) were added to deal with the
underlying business processes. With this extension, visualization tools can
directly represent the concepts of interest for the bank. All these new classes
were defined as subclasses of existing ones in the cartography metamodel (darker
in Fig. \ref{Figure: BusinessCartography}).

The necessary data was provided by the bank in an Microsoft
Excel worksheet. It has been refactored and processed as seen in section \ref{Discovery}.

The most appropriate visualizations for this use case were the Prefuse-based
graph viewers. After analyzing their requests, we used both generic and
weighted graph views (see, in Fig. \ref{Figure: Decision-maker Perspective}, the
Decision-Maker perspective for an example of such a weighted graph), and
predefined a new view definition which, based on the available information,
automatically computes all possible links between all the tools by considering
the formats they export and/or import. Computation was implemented by an ATL
ExtractTools transformation.

By reducing the unforeseen impact of the change, this significantly facilitates
the decisions about which tools to be added/replaced.

%% file: src_uc_eclipse.tex
\subsection{Plug Cartography in an Existing Metamodel: Eclipse B3 Build
	Models}

Nowadays, complex software systems, such as the ones built on top of the Eclipse
platform, are implemented by assembling components coming from different
repositories. Designing build definitions for this kind of systems is complex
since all the dependencies between components must be taken into account.

\begin{figure}\centering
  \includegraphics[width=\linewidth]{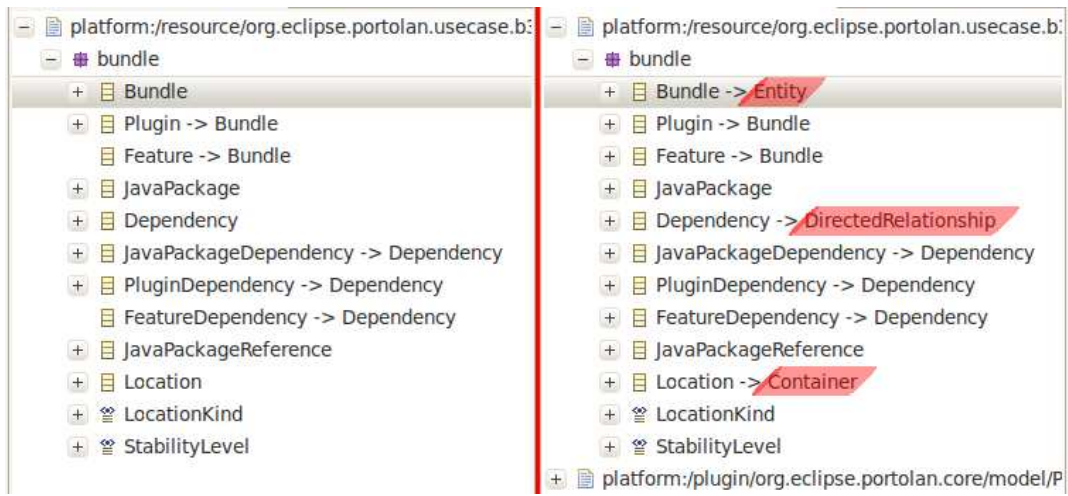}
  \caption{$[$Before$|$After$]$ B3 metamodel}
  \label{Figure: EclipseMMs}
\end{figure}

Eclipse is proposing the new Eclipse B3 build tool for this task. This tool is
a model-based tool and thus it defines a b3 metamodel (Fig. \ref{Figure:
EclipseMMs} on the left) to handle the build definition. Therefore, in order to
offer designers with a  visualization of all components in the build definition,
we can connect the b3 metamodel with our cartography metamodel (Fig.
\ref{Figure: EclipseMMs} on the right) in order to benefit from the
visualization capabilities of Portolan.

\begin{figure}\centering
  \includegraphics[width=\linewidth]{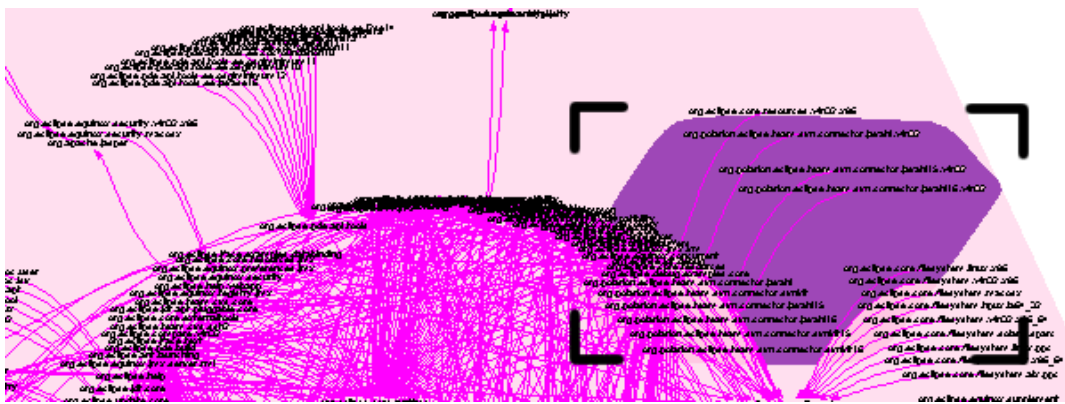}
  \caption{B3 grouped view (excerpt)}
  \label{Figure: B3 Visu2b}
\end{figure}

Once both metamodels are linked, and the underlying models have been translated
via model transformations, a basic graph visualization shows up the complexity
of the plugin dependencies. Using a grouping feature of the viewer, we could
easily detect dependencies which target plug-ins on an inappropriate update
site (colored zone of the graph in Fig. \ref{Figure: B3 Visu2b}).

%% file: src_state_of_art.tex
\section{Related Work}
\label{Related Works section}

Our approach relies on Model-Driven Engineering to provide visualization
capabilities to every business domain that needs it. It offers the ability to
assist in the different tasks required to transform raw user data into the
corresponding graphical views by means of a (predefined set of) injectors,
extractors and model transformations.

Up to now, research on this area has focused on the purely visualization aspects
without considering the whole cartography process.

Most generic visualization tools rely on an underlying technical library such as
Prefuse\cite{heer2005prefuse}, and force users to follow that input format for
the data. This results in adhoc translation mechanisms between the domain data
format and the visualization one. Some examples are the \textit{Bee/Hive}
visualization back-end\cite{reiss2001bee} that relies on OQL and SQL queries;
\textit{Gsee}\cite{favre2001gsee} dedicated to source-code exploration;
\textit{SHriMP}\cite{storey2002shrimp} for hierarchically structured
information. On a higher level, Ma\cite{ma2002visualizing} proposed a tool to
visualize the visualization process itself, allowing the definition of visual
transformations that can then be executed. However, these transformations are
already performed on the visualization itself and thus, are not useful to adapt
the domain data. Our approach provides a more homogeneous framework to develop
reusable transformations.

Some approaches have introduced MDE techniques to deal with visualization
problems at a higher abstraction level. Bull et al.\cite{bull2006architecture}
proposed a framework for view customizations. This framework relied on the
creation of an explicit metamodel for every visualization paradigm or tool
(tree view, graph view, charts, etc). Domokos, Varro and
Varr\'o\cite{domokos2002open} proposed a framework focused on the visualization
of metamodels. Again, while useful, these approaches do not cover the full
cartography process.

A more complex approach is S. Buck et al.\cite{buckl2007generating} that
proposed an automated approach based on model transformations but linked to a
specific and simple visualization metamodel. B. Brodaric and J.
Hastings\cite{brodaric2002object} follow a similar approach but focused on
geographical information systems.
Other domain-specific systems are N. Abdat and Z.
Alimazighi\cite{abdat2008gmtool} with a metamodel with geographic items, F.
Allilaire\cite{allilaire2009towards} for software cartography. Instead, in our
case, the pivot cartography metamodel provides a clear separation of concerns
between both data and visualization, so that designers can easily change one or
the other reusing at least half of the work done.

%% file: src_discussion.tex
\section{Conclusions and further work}
\label{Discussion section}

We have presented our cartography framework to manage the whole chain of tasks
from data capture to visualization. Our approach takes advantage of MDE
techniques to offer an integrated, generic and extensible environment to deal
with graphical display of information for every specific domain. Our framework
has been implemented, as the Portolan tool, on top of the Eclipse platform and
has been validated in three different case studies.

As further work, we plan to enrich the Portolan framework with additional
predefined visualization tools. Some advanced visualization approaches, such as
N. Henry MatrixExplorer\cite{henry2006matrixexplorer} and
NodeTrix\cite{henry2007nodetrix}, could help to deal with huge and dense graphs
and give the users more handles on their data. We will also test Portolan in other
scientific fields to see the benefits of applying our approach on huge amounts
of data. Working on clustering algorithms for models could be needed in this
kind of scenario.